\documentclass[twocolumn,superscriptaddress]{revtex4}
\usepackage[dvips]{graphicx}
\usepackage{amssymb,amsfonts,amsmath}
\usepackage{color}
\usepackage{eurosym}
\usepackage[sort&compress]{natbib}
\usepackage{latexsym}
\usepackage{amssymb}
\usepackage{amsfonts}
\usepackage{amsmath}
\usepackage{theorem}
\usepackage{graphicx}

\newcommand{\be}{\begin{equation}}
\newcommand{\ee}{\end{equation}}
\newcommand{\bd}{\begin{displaymath}}
\newcommand{\ed}{\end{displaymath}}
\newcommand{\BE}{\begin{eqnarray}}
\newcommand{\EE}{\end{eqnarray}}

\newcommand{\bh}{\ensuremath{\mathbf{h}}}

\newcommand{\bx}{\ensuremath{\mathbf{x}}}
\newcommand{\by}{\ensuremath{\mathbf{y}}}
\newcommand{\bz}{\ensuremath{\mathbf{z}}}

\newcommand{\boldpsi}{{\mbox{\boldmath $\psi$}}}
\newcommand{\boldphi}{{\mbox{\boldmath $\varphi$}}}

\newcommand{\avg}[1]{\left\langle{#1}\right\rangle}

\newcommand{\bs}{\bm{s}}

\begin{document}
 \pdfoutput=1
\title{Complex dynamics in learning complicated games}
\author{Tobias Galla}
\email{tobias.galla@manchester.ac.uk}
\address{Theoretical Physics, School of Physics and Astronomy, \\
The University of Manchester, Manchester M13 9PL, United Kingdom}
 \author{J. Doyne Farmer}
 \email{jdf@santafe.edu}
 \address{Santa Fe Institute, 1399 Hyde Park Road, Santa Fe, NM 87501}
 \address{ LUISS Guido Carli, Viale Pola 12, 00198 Roma, Italy}



\begin{abstract}
Game theory is the standard tool used to model strategic interactions in evolutionary biology and social science.  Traditional game theory studies the equilibria of simple games. But is traditional game theory applicable if the game is complicated, and if not, what is?  We investigate this question here, defining a complicated game as one with many possible moves, and therefore many possible payoffs conditional on those moves.  We investigate two-person games in which the players learn based on experience.  By generating games at random we show that under some circumstances the strategies of the two players converge to fixed points, but under others they follow limit cycles or chaotic attractors.  
The dimension of the chaotic attractors can be very high, implying that the dynamics of the strategies are effectively random.   In the chaotic regime the payoffs fluctuate intermittently, showing bursts of rapid change punctuated by periods of quiescence, similar to what is observed in fluid turbulence and financial markets. Our results suggest that such intermittency is a highly generic phenomenon, and that there is a large parameter regime for which complicated strategic interactions generate inherently unpredictable behavior that is best described in the language of dynamical systems theory.
\\ \\
\end{abstract}

\maketitle


Traditional game theory usually gives a good understanding for simple games with a few players, or with only a few possible moves, characterizing the solutions in terms of their equilibria \cite{vonneumann,nash}.  The applicability of this approach is not clear when the game becomes more complicated, for example due to more players or a larger strategy space, which can cause an explosion in the number of possible equilibria \cite{opper,opper2,berg,berg2}.  This is further complicated if the players are not rational and must learn their strategies \cite{ho,camerer,camerer1,fudenberg,young}.  In a few special cases it has been observed that the strategies display complex dynamics and fail to converge to equilibrium solutions \cite{sato,skyrms,hommes}.  Are such games special, or is this typical behavior?  More generally, under what circumstances should we expect that games become so hard to learn that their dynamics fail to converge?  What kind of behavior should we expect and how should we characterize the solutions?

As an example of what we mean compare the games of tic-tac-toe and chess.  Tic-tac-toe is a simple game with only $765$ possible positions and $26,830$ distinct sequences of moves.  Young children easily discover the Nash equilibrium, which results in a draw, at which point the game becomes uninteresting.  In contrast, chess is a complicated game with roughly $10^{47}$ possible positions and $10^{123}$ possible sequences of moves; despite a huge effort, the Nash equilibrium (corresponding to an ideal game) remains unknown.  Equilibrium concepts of game theory are not useful in describing complicated games such as chess or go (which has an even larger game tree with roughly $10^{360}$ possible sequences of moves).  An example that is even closer to what we have in mind here is investing in financial markets, which is a non-zero sum game where players can choose between thousands of assets and a rich set of possible strategies.

Here we show that if the players use a standard approach to learning, for complicated games there is a large parameter regime in which one should expect complex dynamics.  By this we mean that the players never converge to a fixed strategy.  Instead their strategies continually vary as each player responds to past conditions and attempts to do better than the other players.  The trajectories in the strategy space display high-dimensional chaos, suggesting that for most intents and purposes the behavior is essentially random, and the future evolution is inherently unpredictable.

\section{\bf Model}
\begin{figure*}[t!!!]
\includegraphics[scale=0.4]{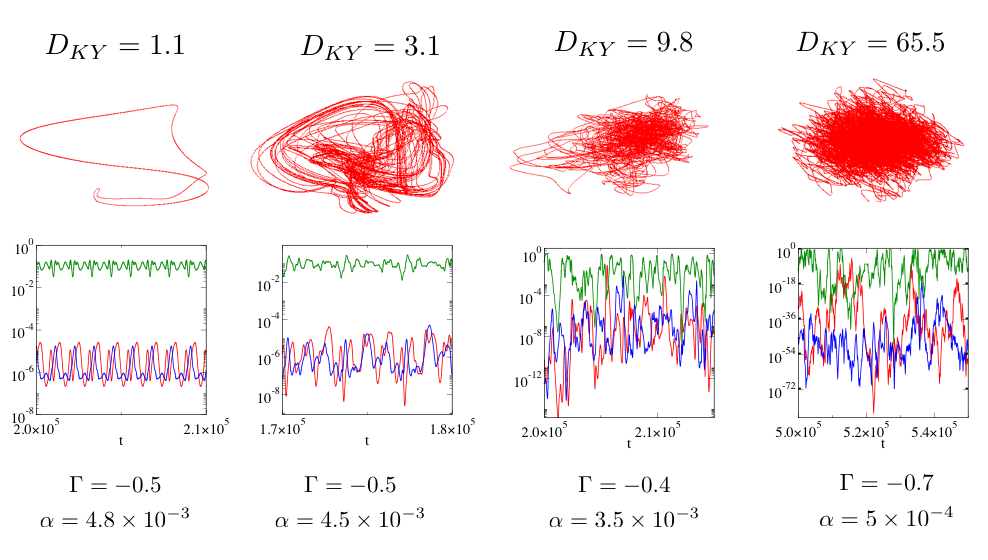} 
\caption{An illustration of complex learning dynamics, depicted in terms of trajectories of the strategy ${\bf x}^\mu(t)$ for different parameters ($\beta=0.01$).   In (a) the attractor is a limit cycle, whereas (b-d) are chaotic attractors of increasingly high dimension. There are $N = 50$ possible moves; the upper panel shows an arbitrary three-dimensional projection of each attractor in the $98$-dimensional phase space, and lower panels show the strategy as a function of time for the corresponding three coordinates.  For clarity we use logarithmic scale.   As the dimension of the attractor increases, so does the range of $x^\mu_i$.   For the highest dimensional case a given move has occasional bursts where it is highly probable, and long periods where it is extremely improbable (as low as $10^{-72}$).}
\label{fig:fig1}
\end{figure*}

To address the questions raised above we study two-player games.    For convenience call the two players Alice and Bob.  At each time step $t$ player $\mu \in \mbox{\{Alice = A, Bob = B\}}$ chooses between one of $N$ possible moves, picking the $i^{th}$ move with frequency $x_i ^\mu (t)$, where $i=1, \ldots, N$.  The frequency vector ${\bf x}^\mu (t)=(x_1^\mu,\dots,x_N^\mu)$ is the strategy of player $\mu$.  If Alice plays $i$ and Bob plays $j$, Alice receives receives payoff $\Pi_{ij}^A$ and Bob receives payoff $\Pi_{ji}^B$.  

We assume that the players learn their strategies ${\bf x}^\mu$ via a form of reinforcement learning called {\it experience weighted attraction}.  This has been extensively studied by experimental economists who have shown that it provides a reasonable approximation for how real people learn in games \cite{ho,camerer1,camerer}.  Actions that have proved to be successful in the past are played more frequently and moves that have been less successful are played less frequently.  To be more specific, the probability of a given move is 
\begin{equation}\label{eq:x}
x_i^\mu (t) = \frac{e^{\beta Q_i^\mu (t)}}{\sum_k e^{\beta Q_k^\mu (t)}},
\end{equation}
where $Q_i^\mu$ is called the  {\it attraction} for player $i$ to strategy $\mu$.  In the special case of experience weighted attraction that we use here, Alice's attractions are updated according to
\begin{equation}\label{eq:q}
Q_i^A (t+1) = (1 - \alpha) Q_i^A (t) + \sum_j \Pi_{ij}^A x_j^B
\end{equation}
and similarly for Bob with $A$ and $B$ interchanged. 

The dynamics for updating the strategies ${\bf x}^\mu$ of the two players are completely deterministic.  This approximates the situation in which the players vary their strategies slowly in comparison to the timescale on which they play the game.  

The key parameters that characterize the learning strategy are $\alpha$ and $\beta$.  The parameter $\beta$ is called the intensity of choice; when $\beta$ is large a small historical advantage for a given move causes that move to be very probable, and when $\beta = 0$ all moves are equally likely. The parameter $\alpha$ specifies the memory in the learning; when $\alpha$ = 1 there is no memory of previous learning steps, and when $\alpha = 0$ all learning steps are remembered and are given equal weight, regardless of how far in the past.  The case $\alpha = 0$ corresponds to the much-studied replicator dynamics used to describe evolutionary processes in population biology \cite{sato2,nowak,sigmund}.

We choose games at random by drawing the elements of the payoff matrices $\Pi_{ij}^\mu$ from a normal distribution \cite{may,opper,opper2,berg,berg2}.  The mean and the covariance are chosen so that 
$E[\Pi_{ij}^\mu] = 0$,  $E[(\Pi^\mu_{ij})^2] = 1/N$, and $E[\Pi_{ij}^A \Pi_{ji}^B] = \Gamma/N$, where $E[x]$ denotes the average of $x$. 
The variable $\Gamma$ is a crucial parameter which measures the deviation from a zero-sum game.  When $\Gamma = -1$ the game is zero sum, i.e. the amount Alice wins is equal to the amount Bob loses, whereas when $\Gamma = 0$ their payoffs are uncorrelated.

\section{\bf Results}
We simulate randomly constructed games with $N = 50$ possible moves, corresponding to a $98$ dimensional state space (there are two $50$ dimensional strategy vectors and two probability constraints).  The behavior observed depends on the parameters.  In some cases we see stable learning dynamics, in which the strategies ${\bf x}^\mu$ of both players converge on a fixed point.  For a large section of the parameter space, however, the strategies converge to a more complicated orbit, either a limit cycle or a chaotic attractor.  We characterize the local stability properties of the attractors by numerically computing the Lyapunov exponents $\lambda_i$, $i = 1, \ldots, 2N-2$,  which quantify the rate of expansion or contraction of nearby points in the state space.  The Lyapunov exponents also determine the Lyapunov dimension $D$, which measures the number of degrees of freedom of the motion on the attractor.  

We give several examples of the observed learning dynamics at different parameter values in Fig.~\ref{fig:fig1}.  These include a limit cycle and chaotic attractors of varying dimensionality.   There can also be long transients in which the trajectory follows a complicated orbit for a long time and then suddenly collapses into a fixed point.  In general the behavior observed depends on the random draws of the  payoff matrices $\Pi_{ij}^\mu$, but as we move away from the stability boundary, for a given set of parameters we observe fairly consistent behavior .

\begin{figure}[t!]
\includegraphics[scale=0.325]{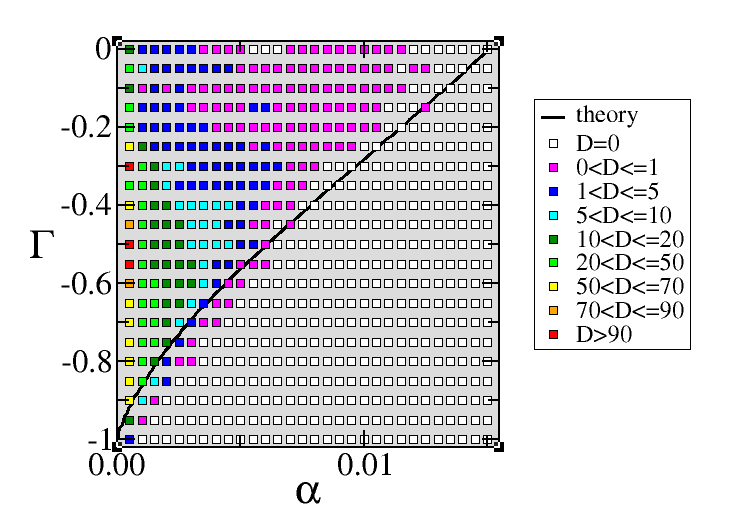} 
\caption{Stability diagram showing where stable vs. chaotic learning is likely ($\beta=0.01$). The solid line is the stability boundary estimated using path-integral methods.  The coloured squares are from simulations of the learning dynamics and represent the typical dimension of the attractor (averaged over $10$ or more independent payoff matrices per data point).}
\label{fig:fig2}
\end{figure}

Simulating games at many different parameter values reveals the stability diagram given in Fig. \ref{fig:fig2}.   Roughly speaking we find that the dynamics are stable \cite{f1} when $\Gamma \approx -1$ (zero sum games) and $\alpha$ is large (short memory), i.e. in the lower right of the diagram, and unstable when $\Gamma \approx 0$ (uncorrelated payoffs) and $\alpha$ is small (long memory), i.e. in the upper left.  Interestingly, for reasons that we do not understand the highest dimensional behavior is observed when the payoffs are moderately anti-correlated ($\Gamma\approx -0.6$) and when players have long-memory ($\alpha\approx 0$). 

In order to make the problem analytically tractable we have made specific choices in the parameters for experience weighted attraction (EWA).  Comparison with behavioral experiments modeled with EWA as reported in \cite{ho} shows that the particular form we are using here is roughly within the range observed in real experiments.  Values for memory-loss parameters and intensity of choice reported from experiments suggest that real-world decision making may well operate near or in the chaotic phase (see Supplementary Information). Most experimental data is limited to low-dimensional games however, whereas here we study games with a large number of possible moves.  A good example where high dimensional chaotic behavior is likely is in financial markets, where there are a huge number of possible moves and learning times are measured in years.  High dimensional chaotic behavior can be effectively indistinguishable from noise.

A good approximation of the boundary between the stable and unstable regions of the parameter space can be computed analytically using techniques from statistical physics. We use path-integral methods from the theory of disordered systems \cite{dedominicis,opper} to compute the stability in the limit of infinite payoff matrices, $N\to\infty$.  We do this in a continuous-time limit where, for fixed $\Gamma$,  stability then depends only on the ratio $\alpha/\beta$ (see Supplementary Information).

We have simulated games for various values of $N$.  If $D > 0$ at small $N$, the dimension $D$ tends to increase with $N$.  At this stage we have been unable to tell whether $D$ reaches a finite limit or grows without bound as $N \to \infty$.

\begin{figure}[t!!!]
\includegraphics[scale=0.4]{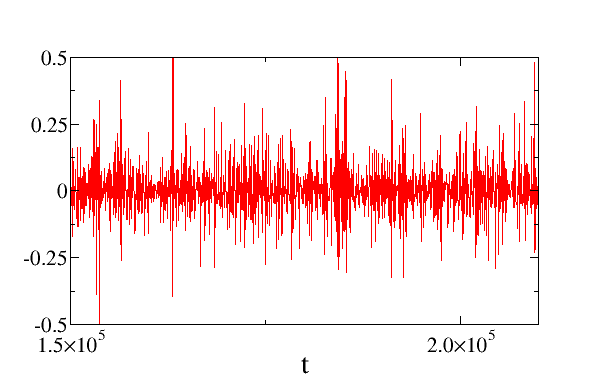} 
\caption{{\em Chaotic dynamics display clustered volatility}. We plot the difference of payoffs on successive time steps for case (c) in Fig. \ref{fig:fig1}.  The amplitude of the fluctuations increases with the dimension of the attractor.}
\label{fig:fig3}
\end{figure}

An interesting property of this system is the time dependence of the received payoffs.  As shown in Fig.~\ref{fig:fig3}, when the dynamics are chaotic the total payoff to all the players varies, with intermittent bursts of large fluctuations punctuated by relative quiescence.  This is observed, although to varying degrees, throughout the chaotic part of the parameter space.  There is a strong resemblance to the clustered volatility observed in financial markets, which in turn resembles the intermittency of fluid turbulence \cite{ghashghaie,hommes}.  We also observe heavy tails in the distribution of the fluctuations, as described in more detail in the Supplementary Information.  This suggests that these properties, which have received a great deal of attention in studies of financial markets, may occur simply because they are generic properties of complicated games \cite{f2}.

 \section{\bf Why is dimensionality relevant?}

The dimensionality $D$ is relevant to this problem because high dimensionality suggests that failure to converge to a fixed point is independent of the learning algorithm, i.e. the game is intrinsically hard to learn.  The fact that the equilibria of a game are unlearnable with any particular learning algorithm, such as reinforcement learning, does not imply that learning is not possible with some other learning algorithm. For example, if the learning dynamics settles into a limit cycle or a low dimensional attractor, a careful observer could collect data and make better predictions about the other player using the method of analogues \cite{lorenz}, or refinements based on local approximation \cite{farmer}.  If the dimension of the chaotic attractor is too high, however, the curse of dimensionality makes this impossible with any reasonable amount of data \cite{farmer}. This suggests that there exists no learning algorithm that can provide an improvement when learning must occur inductively based on past data.  The observation of high-dimensional dynamics here leads us to conjecture that there are some games that are inherently unlearnable, in the sense that any learning algorithms will inevitably result in high-dimensional chaotic learning dynamics (See also Sato et al. \cite{sato}).

Our work here makes it possible to predict {\it a priori} the qualitative properties of the learning dynamics of any given complicated two player game under reinforcement learning.  This is because the payoff matrix of any given game is a possible draw from an ensemble of random games.  One can make a good estimate of the stability properties of the learning dynamics by locating the game and the learning parameters in the stability diagram of Fig.~\ref{fig:fig2}.  We have shown that a key property of a game is its ``zero-sumness", characterized by $\Gamma$.  Games become harder to learn (in the sense that the strategies do not converge) when they are non-zero-sum, particularly if the players use learning algorithms with long memory.   This analysis can potentially be extended to multiplayer games, games on networks, alternative learning algorithms, etc. 

Our results suggest that under many circumstances it is more useful to abandon the tools of classic game theory in favor of those of dynamical systems.  It also suggests that many behaviors that have attracted considerable interest, such as clustered volatility in financial markets, may simply be specific examples of a highly generic phenomenon, and should be expected to occur in a wide variety of different situations.
 
 \section*{Acknowledgements} We would like to thank the National Science Foundation for grant 0624351. TG is supported by an RCUK Fellowship (reference EP/E500048/1) and would like to thank EPSRC (UK) for support (grants EP/I019200/1 and EP/I005765/1). We would also like to thank Yuzuru Sato and Nathan Collins for useful discussions.

\onecolumngrid
\appendix
 \section{Experience weighted attraction learning}
\subsection{General definitions for multi-player games}
We here briefly describe the experience weighted attraction learning (EWA) model put forward in \cite{camerer,camerer1,camerer2}. Consider a game played by $p$ players, who each choose from a set of $N$ actions (pure strategies) at each time step\footnote{Generalisation to games in which different players have different numbers of actions at their disposal is straightforward.}.  In the EWA model the probability for player $\mu\in\{1,\dots,p\}$ to choose action $i\in\{1,\dots,N\}$ at time $t$ is 
\be\label{eq:xdef}
x^\mu_i(t)=\frac{e^{\beta Q^\mu_i(t)}}{\sum_k e^{\beta Q^\mu_k(t)}},
\ee
where the $\{Q_i^\mu\}$ are referred to as attractions or propensities\footnote{In \cite{camerer2} the attractions are denoted by $A_\mu^i(t)$,  the components of the mixed strategies by $P_\mu^i$.}. The basic idea is that $Q^\mu_i(t)$ gives the ``attraction" of player $\mu$ to action $i$ at time $t$, based on how successful strategy $i$ has been in the past. The model parameter $\beta\geq 0$ is called the {\it intensity of choice}.  For $\beta=0$ players pick actions with equal probability (i.e. they play completely at random), and for $\beta\to\infty$ each player's choice is deterministic, i.e. that player will always choose the same action for given values of $Q_i^\mu$, namely the action with the highest attraction.

The update rule for the attractions $\{Q^\mu_i\}$ in the EWA model reads \cite{camerer,camerer1, camerer2}
\be\label{eq:qupdate0}
Q^\mu_i(t+1)=\frac{\phi {\cal N}(t) Q^\mu_i(t)+[\delta+(1-\delta)I(i,s_\mu(t)] \Pi^\mu(i,\bs_{-\mu}(t))}{{\cal N}(t)},
\ee
where the quantity ${\cal N}(t)$ is updated according to
\be\label{eq:nupdate}
{\cal N}(t+1)=\phi(1-\kappa){\cal N}(t)+1,
\ee
see \cite{camerer1}.  The notation is explained below:
\begin{itemize}
\item 
We write $s_\mu(t)\in\{1,\dots,N\}$ for the action player $\mu$ takes at time step $t$ in a given realization of the dynamics.  The notation $-\mu$ labels all players other than $\mu$, i.e. $-\mu$ is the set $\{1,\dots,p\}\backslash \{\mu\}$.  The notation $\bs_{-\mu}(t)$ indicates the set of actions that the opponents of player $\mu$ took in a given round.  Thus $\bs_{-\mu}\in\{1,\dots,N\}^{p-1}$ is a $(p-1)$-component vector, each component of which is one of the possible actions.
\item 
For a given $\bs_{-\mu}(t)\in\{1,\dots,N\}^{p-1}$ the quantity $\Pi^\mu(i,\bs_{-\mu}(t))$ is a payoff matrix element, and indicates the payoff player $\mu$ receives when playing pure strategy $i$ and facing the actions $\bs_{-\mu}(t)$ of the other agents at time $t$.  
\item 
The variable $\cal{N}$ indicates a weight factor.  An initial condition needs to be specified, which is then updated according to relation (\ref{eq:nupdate}).  When $\phi(1 - \kappa) = 1$, this is just the number of times the game has been played.  In this case, since $\cal{N}$ cancels in the first term in the numerator on the RHS of Eq. (\ref{eq:qupdate0}), and since it divides the second term, as time goes on the influence of the updates becomes smaller and smaller, i.e. past moves have more weight than recent moves and the behavior becomes ``set".  
\item The notation $I(\cdot,\cdot)$ stands for the indicator function (also called the Kronecker delta), i.e. $I(a,b)= 1$ if $a=b$ and $I(a,b)=0$ otherwise.
\item The parameter $\delta$ specifies the relative weighting given to strategies that are played vs. those that are not played. In the case of $\delta=1$ players update all attractions $Q^\mu_i$ in every round, irrespective of what actions they actually took. The choice $\delta=0$ corresponds to a case where only the scores of strategies that are actually used in a given round are updated after that round.
\item The parameter $\kappa$  interpolates between average reinforcement learning ($\kappa=0$) and cumulative reinforcement learning ($\kappa=1$), see \cite{camerer,camerer1,camerer2}; we have ${\cal N}(t)=1$ for all $t$ if $\kappa=1$, the attractions $Q_i^\mu$ then represent the cumulative outcome of all past play (depending on the choice of $\phi$ potentially discounted over time), for $\kappa=0$ the normalisation factor ${\cal N}(t)$ grows with time.
\item The parameter $\phi$ specifies the weight of outcomes of play in the distant past relative to more recent iterations. If $\kappa=1$ and $\phi=1$ all past experience carries equal weight, no matter how much time has elapsed, for $\phi=0$ only the most recent round affects the players' future decisions. Intermediate values of $\phi$ correspond to exponential discounting.
\end{itemize}
\subsection{The specific case that we study here}
There are many parameters within the formalism of experience weighted attraction, and it is beyond the scope of this paper to investigate all of the possible cases.  We thus restrict ourselves to a particular case that is both analytically tractable and reasonably close to how real people play.

We first assume that Eq. (\ref{eq:nupdate}) reaches a fixed point $\cal{ N}^*$ in the long-run.  Letting ${\cal{N}}(t+1) = {\cal{N}}(t)={\cal N}^*$ gives
\be
{\cal N}^*=\frac{1}{1-\phi(1-\kappa)}.
\ee
The update rule then simplifies to 
\be
Q^\mu_i(t+1)=\phi Q^\mu_i(t)+(1-\phi(1-\kappa))[\delta+(1-\delta)I(i,\bs_\mu(t)] \Pi^\mu(i,\bs_{-\mu}(t)).
\ee

We will focus here on the case $\delta=1$, i.e. all strategy scores are updated in every iteration. Then we have
\be\label{eq:qupd2}
Q^\mu_i(t+1)=\phi Q^\mu_i(t)+(1-\phi(1-\kappa)) \Pi^\mu(i,\bs_{-\mu}(t)).
\ee
Focusing on cumulative re-inforcement learning \cite{camerer,camerer1,camerer2}, i.e. the case $\kappa=1$, and replacing $\phi\rightarrow 1-\alpha$ for later convenience,  we have\footnote{For average re-inforcement learning \cite{camerer,camerer1,camerer2}), i.e. $\kappa=0$, one has
$Q^\mu_i(t+1)=\phi Q^\mu_i(t)+(1-\phi)\Pi^\mu(i,\bs_{-\mu}(t))$.  Assuming $\phi\neq 1$ this can be re-scaled to give $\widetilde {Q^\mu_i}(t+1)=\phi\widetilde{Q^\mu_i}(t)+\Pi^\mu(i,s_{-\mu}(t))$, where  $\widetilde {Q^\mu_i}=Q^\mu_i/(1-\phi)$. This update rule is of the type (\ref{eq:scupd}) as well, the rescaling of the propensities amounts to a rescaling of the model parameter $\beta$ in Eq. (\ref{eq:xdef}).} 
\be\label{eq:scupd}
Q^\mu_i(t+1)=(1-\alpha) Q^\mu_i(t)+\Pi^\mu(i,\bs_{-\mu}(t)).
\ee
Eq. (\ref{eq:scupd}) is the learning rule used in \cite{sato,sato2}. The parameter $\alpha$ describes memory loss. For $\alpha=0$ past payoffs are not discounted, and the memory of players covers the full history of play. For $0<\alpha<1$ past payoffs are taken into account with exponentially decreasing weights.
 
To summarise, the learning model we investigate is defined by Eq. (\ref{eq:xdef}) together with Eq. (\ref{eq:scupd}):
\begin{itemize}
\item Eq. (\ref{eq:xdef}) specifies how a given player $\mu\in\{1,\dots,p\}$ translates his or her set of attractions $Q^\mu_i$, $i\in\{1,\dots,N\}$, into a mixed strategy $(x_1^\mu,\dots,x_N^\mu)$. He or she will choose action $\mu\in\{1,\dots,N\}$ with the probabilities defined by Eq. (\ref{eq:xdef}).  
\item Eq. (\ref{eq:scupd}) specifies how the attractions are updated from time step $t$ to $t+1$ once all players have chosen their actions in time step $t$.
\end{itemize}
The correspondence with the EWA model of \cite{camerer1} can be summarized as follows:
\BE
\mbox{model of Camerer et al.} &\leftrightarrow&\mbox{notation in present work} \nonumber \\
P_\mu^i ~~~~&\leftrightarrow& ~~~~ x^\mu_i, \nonumber \\
A_\mu^i ~~~~&\leftrightarrow& ~~~~Q^\mu_i, \nonumber \\
\phi ~~~~&\leftrightarrow& ~~~~1-\alpha.
\EE

\subsection{Relation to experimental data}

Parameters of the EWA learning model were fit to real-world data in \cite{camerer2}, see in particular Table 4. This table shows that there is substantial variation in the parameters that provide a best fit to the data across different games.  While we have chosen parameters that were tractable for the theoretical calculations that follow, comparison to their experimental results indicates that these values are fairly reasonable.  For example, they find values of the parameter $\kappa$ in the range $0.15$-$0.99$; we fix $\kappa=1$. The model parameter $\delta$ obtained from experimental data varies from $\delta=0$ to $\delta=0.94$, suggesting that there is no clear conclusion on whether or not players use forgone payoffs in their adaptation; we fix $\delta=1$, i.e. the propensities of all strategies are updated at every step. 

The most interesting model parameters from the point of view of our analysis are the memory-loss parameter $\alpha$ and the intensity of choice $\beta$. For a fixed game, these parameters largely determine whether or not one should expect convergence or chaotic motion. More precisely the ratio $\alpha/\beta$ is the crucial indicator for the onset of chaos, as explained above. Pooled data from \cite{camerer2} suggests a ratio of $\alpha/\beta\approx 0.03$ (but again with considerable variation across games). Depending on the character of the game (zero-sum or not) this can position such experiments inside the chaotic phase, see Fig. \ref{fig:suppfig2}. It is important to keep in mind, however, that the games used in the experiments of \cite{camerer2} are low-dimensional, in the sense that each player has the choice only between a small number of moves. Care needs to be taken when extrapolating results for high-dimensional random games to these cases.  Nonetheless, if one assumes that parameters would not dramatically change in moving from simple games to complicated games, then the data and model fitting of \cite{camerer1}, taken together with our results, suggests that real-world learning in non-zero sum games may well operate in or near the chaotic phase. 

\subsection{Adiabatic limit and deterministic learning}
The update of Eq. (\ref{eq:scupd}) is intrinsically stochastic, as the $(p-1)$-component action vector $\bs_{-\mu}(t)$ at time $t$ is drawn according to the mixed strategy profiles of the $p-1$ opponents player $\mu$ is facing. More precisely, player $\mu$ will face a specific realization of the actions $\bs_{-\mu}=(s_1,\dots,s_{\mu-1},s_{\mu+1},\dots,s_p)\in\{1,\dots,N\}^{p-1}$ of all the other players with probability 
\be
\bx^{-\mu}_{\bs_{-\mu}}=\prod_{\nu\neq\mu}x^\nu_{s_\nu}.
\ee

In order to simplify the problem we follow \cite{sato,sato2} and consider an adiabatic limit of this process. This corresponds to averaging over batches of a large (infinite) number of rounds between two adaptation steps i.e. to the replacement
\be
\Pi^\mu(i,\bs_{-\mu}(t)) \longrightarrow \overline{\Pi^\mu_i}(t):=\sum_{\bs_{-\mu}}  \Pi^\mu(i,\bs_{-\mu})\bx^{-\mu}_{\bs_{-\mu}}(t).
\ee
in Eq. (\ref{eq:scupd}).
The sum over $\bs_{-\mu}$ here runs over all elements of $\{1,\dots,N\}^{p-1}$. We have introduced the notation $\overline{\Pi^\mu_i}(t)$ to indicate that the right-hand-side is the mean of the left-hand-side, i.e. $\overline {\Pi_i^\mu}(t)$ is the {\em expected} payoff for player $\mu$ if she chooses to play action $i$ and given her opponents' mixed strategy profiles at time $t$. In this sense the adiabatic limit can be understood as describing the dynamics on expectation. Fluctuation effects induced by the stochastic choice of pure actions by the players are here neglected, see however \cite{gallaprl,realpe,gallanoise} for systematic studies of noisy learning in simple games.
\\

The equation for updating the attractions becomes
\be
Q_i^\mu(t+1)=(1-\alpha) Q_i^\mu(t)+\overline{\Pi^\mu_i}(t).
\ee

Taking into account Eq. (\ref{eq:xdef}) the learning process can then be described by the following deterministic map
\be\label{eq:mapmulti}
x^\mu_i(t+1)=\frac{x^\mu_i(t)^{1-\alpha}e^{\beta\overline{\Pi^\mu_i}(t)}}{\sum_k x^\mu_k(t)^{1-\alpha}e^{\beta\overline{\Pi^\mu_k}(t)}}.
\ee
Here, each player chooses between $N$ actions, $i=1,\dots,N$, and there are $p$ players, so we have $p\times N$ variables $\{x^\mu_i\}$ in total. These variables satisfy the constraints $\sum_i x^\mu_i(t)=1$ at all times $t$ for all $\mu=1,\dots,p$. Eq. (\ref{eq:mapmulti}) therefore defines a map in a $p\times(N-1)$-dimensional phase space.

 \section{Details of the two-player learning model}

\subsection{Definition of the dynamics}
While the previous sections described learning in a general $p$-player game, we will now restrict the further discussion to the case $p=2$, i.e. to two-player games. The two players are Alice (A) and Bob (B). Each of them has $N$ strategies to choose from. 
Eqs. (\ref{eq:scupd}) then read
\BE
Q^A_i(t+1)&=&(1-\alpha) Q^A_i(t)+\Pi^A(i,s_{B}(t)),\nonumber \\\
Q^B_i(t+1)&=&(1-\alpha) Q^B_i(t)+\Pi^B(i,s_{A}(t)).
\EE
\underline{Simplication of notation:}\\
We will write $x_i(t)$ for the probability with which Alice uses action $i$ at time $t$, and similarly $y_i(t)$ is the probability with which Bob plays action $i$ at that time\footnote{It is here important to remember that, for notational convenience, we enumerate Alice's strategies by $i\in\{1,\dots,N\}$ and similarly for Bob. We do not imply that Alice's action with a given label $i$ is identical to Bob's action with the same label. The games we considering are general asymmetric two-player $N$-action games.}. For simplicity we will change the notation by letting $a_{ij}$ be the payoff Alice receives when she plays action $i$ and when Bob plays action $j$. The payoff for Bob in this situation will be $b_{ji}$. The two $N\times N$ matrices $(a_{ij})$ and $(b_{ij})$, with $i,j\in\{1,\dots,N\}$ then define an asymmetric two-player game, in which each player has $N$ pure strategies to choose from. Taking the deterministic limit, as described above, the update rules for the attractions now read
\BE
Q^A_i(t+1)&=&(1-\alpha) Q^A_i(t)+\sum_j a_{ij} y_j(t)\nonumber \\\
Q^B_i(t+1)&=&(1-\alpha) Q^B_i(t)+\sum_j b_{ij} x_j(t),
\EE
and the map of strategy updates is given by
\be\label{eq:map2}
x_i(t+1)=\frac{x_i(t)^{1-\alpha}e^{\beta\sum_j a_{ij} y_j(t)}}{\sum_k x_k(t)^{1-\alpha}e^{\beta\sum_j a_{kj} y_j(t)}}, ~~~~~~ y_i(t+1)=\frac{y_i(t)^{1-\alpha}e^{\beta\sum_j b_{ij} x_j(t)}}{\sum_k y_k(t)^{1-\alpha}e^{\beta\sum_j b_{kj} x_j(t)}}.
\ee
\subsection{Relation between discrete-time dynamics and continuous-time Sato-Crutchfield equations}
Chaotic motion in learning dynamics of the above type has previously been reported for relatively low-dimensional games in \cite{sato,sato2}. These studies were carried out for continuous-time processes, and it is therefore useful to elaborate on the relation of the discrete-time map defined by Eq. (\ref{eq:map2}) and the continuous-time dynamics of Sato et al\footnote{Chaotic motion in discrete-time evolutionary dynamics of low-dimensional games has recently been investigated in \cite{vilone}.}.

The discrete-time dynamics of Eq. (\ref{eq:map2}) can be written as
\BE
x_i(t+1)&=&\frac{x_i(t)^{1-\alpha}e^{\beta\sum_j a_{ij}y_j(t)}}{Z_x(t)}, \nonumber \\
y_j(t+1)&=&\frac{y_j(t)^{1-\alpha}e^{\beta\sum_i b_{ji}x_i(t)}}{Z_y(t)}, \label{eq:mapxyz}
\EE
where we define the normalisation factors
\be
Z_x(t)=\sum_k x_k(t)^{1-\alpha}e^{\beta\sum_j a_{kj}y_j(t)}, ~~~~ Z_y(t)=\sum_k y_k(t)^{1-\alpha}e^{\beta\sum_j b_{kj}x_j(t)}.
\ee
The continuous-time Sato-Crutchfield dynamics on the other hand is given by
\BE
\dot x_i=x_i\left(\sum_j a_{ij}y_j-\alpha'\ln x_i-Z_x'\right), \nonumber \\
\dot y_j=y_j\left(\sum_i b_{ji}x_i-\alpha'\ln y_j-Z_y'\right) \label{eq:cont},
\EE
see \cite{sato,sato2} for details. The parameter $\alpha'$ indicates memory loss in this continuous dynamics, and it is hence analogous to the parameter $\alpha$ in the above map (\ref{eq:mapxyz}). We will detail the relation between $\alpha$ and $\alpha'$ further below. Similarly, the role of $Z_x'$ and $Z_y'$ in Eqs. (\ref{eq:cont}) is to enforce the normalisation $\sum_i x_i=\sum_i y_i=1$ at all times\footnote{Dashed quantities $\alpha', Z_x', Z_y'$ refer to the continuous-time dynamics. Their counterparts without dashes are for the discrete-time process.}. These quantities can be thought of as Lagrange multipliers, they can be expressed explicitly as
\be
Z_x'=\sum_i x_i\left(\sum_j a_{ij}y_j-\alpha' \ln x_i\right), ~~~ Z_y'=\sum_i y_i\left(\sum_j b_{ij}x_j-\alpha' \ln y_i\right).
\ee
Similar to what is the case for $\alpha$ and $\alpha'$ there is a close relation between $Z_x$ and $Z_x'$ and between $Z_y$ and $Z_y'$ respectively. This will be explained in more detail below.
\vspace{1em}

\underline{Limit of small $\beta$:}\\
In order to relate the discrete-time update rule to the continuous-time Sato-Crutchfield dynamics we consider the limit $\beta\ll1$ in Eq. (\ref{eq:mapxyz}). One first writes
\be
\ln x_i(t+1)=(1-\alpha)\ln x_i(t)+\beta\sum_j a_{ij} y_j(t)-\ln Z_x(t),
\ee
and similarly for the second equation of (\ref{eq:mapxyz}). This is valid for all $\beta$, and can be re-arranged to give
\be
\frac{\ln x_i(t+1)-\ln x_i(t)}{\beta}=-\frac{\alpha}{\beta}\ln x_i(t)+\sum_j a_{ij} y_j(t)-Z_x'(t)
\ee
where $Z_x'(t)=\ln Z_x(t)/\beta$. In the limit $\beta\to 0$, fixing the ratio $\alpha/\beta$ during the limiting procedure, and upon appropriate re-scaling of time, this turns into
\be
\frac{d}{dt} \ln x_i(t)=-\frac{\alpha}{\beta}\ln x_i(t)+\sum_j a_{ij} y_j(t)-Z_x'(t),
\ee
i.e.
\be
\dot x_i(t)=x_i(t)\left(-\frac{\alpha}{\beta}\ln x_i(t)+\sum_j a_{ij} y_j(t)-Z_x'(t)\right),
\ee
which is exactly the first equation of the continuous-time dynamics (\ref{eq:cont}), with the replacement $\alpha'=\lim_{\beta\to 0}\alpha/\beta$. A similar argument can be made for the dynamics of $y_i(t)$.

We conclude that the small-$\beta$ limit of the discrete-time dynamics at memory-loss parameter $\alpha$ leads to the continuous-time Sato-Crutchfield dynamics with memory-loss parameter $\alpha'=\alpha/\beta$, after a re-scaling of time.
\vspace{1em}

\underline{Relation of fixed-points}\\
For any choice of $\beta$ the fixed points of the equations
\BE
x_i(t+1)&=&\frac{x_i(t)^{1-\alpha}e^{\beta\sum_j a_{ij}y_j(t)}}{Z_x(t)}, \nonumber \\
y_i(t+1)&=&\frac{y_i(t)^{1-\alpha}e^{\beta\sum_j b_{ij}x_j(t)}}{Z_y(t)} 
\EE
fulfill
\be
\ln x_i^*=(1-\alpha)\ln x_i^*+\beta\sum_j a_{ij} y_j^*-\ln Z_x^*,
\ee
with a similar equation for $y_j^*$. Asterisks here indicate quantities evaluated at the fixed point. We have here assumed that fixed points lie in the interior of the strategy simplex, i.e. that $x_i^*>0$ and $y_i^*>0$ for all $i$.  

These fixed-point conditions can be reduced to
\BE
-\frac{\alpha}{\beta}\ln x_i^*+\sum_j a_{ij} y_j^*-Z_x'^*&=&0 \nonumber \\
-\frac{\alpha}{\beta}\ln y_j^*+\sum_i b_{ji} x_i^*-Z_y'^*&=&0,
\EE
which reproduces the fixed-point condition of the continuous dynamics, see Eqs. (\ref{eq:cont}).
\\

 \underline{Summary:}
 \begin{enumerate}
 \item Up to a re-scaling of time the small-$\beta$ limit of the discrete-time dynamics at parameters $\alpha$, $\beta$ corresponds to the continuous-time Sato-Crutchfield dynamics at parameter $\alpha'=\alpha/\beta$.
 \item For any choice of $\beta$ the fixed points of the map at parameters $\alpha,\beta$ are identical to those of the continuous-time dynamics at $\alpha'=\alpha/\beta$.
 \item Provided a fixed point of the map exists, its components only depend on the ratio $\alpha/\beta$.
 \end{enumerate}
In the following we will use the notation $r=\beta/\alpha$ to denote the relevant control parameter of the continuous dynamics. Given that $\alpha$ can be viewed as a damping parameter, and $\beta$ as a forcing term, the ratio $r=\beta/\alpha$ plays a role similar to that of a Reynolds number in fluid dynamics \cite{satofarmer}.
\subsection{Large random two-player games}
We will now consider the case of large random games. To this end we will follow the standard spin-glass conventions \cite{parisimezardvirasoro, opper,opper2,opper3} and focus on payoff matrices with elements drawn from Gaussian distributions. These distributions are fully characterized by their first and second moments. Specifically we will choose the payoff matrix elements $\{a_{ij},b_{ij}\}$ such that
\BE
{\mathbb E}[a_{ij}]&=&{\mathbb E}[b_{ij}]=0, \nonumber \\
{\mathbb E}[(a_{ij})^2]&=&{\mathbb E}[(b_{ij})^2]=\frac{1}{N}, \nonumber \\
{\mathbb E}[a_{ij}b_{ji}]&=&\frac{\Gamma}{N}  \label{eq:corr}
\EE
for all pairs $i,j\in\{1,\dots,N\}$. The significance of the parameter $\Gamma$ will be explained below. The notation ${\mathbb E}[\cdots]$ denotes the average over the distribution of payoff matrices. Every single element of the payoff bi-matrix is a Gaussian random variable of mean zero. It is important to stress that while the payoff matrices are drawn at random at the beginning, they remain fixed during the time evolution of the dynamics. In the language of spin glass theory \cite{parisimezardvirasoro} they constitute the {\em quenched disorder} of the problem. The factors of $1/N$ in Eqs. (\ref{eq:corr}) indicate that each payoff matrix element is of magnitude $1/\sqrt{N}$. This scaling with $N$ is standard in spin glass theory, and chosen to ensure a non-trivial thermodynamic limit, $N\to \infty$, as explained below. We point out that payoff matrix elements occur in the learning process of Eqs. (\ref{eq:mapxyz}) only in combinations of the type $\beta a_{ij}$ and $\beta b_{ij}$. The choice of scaling of the payoff matrices is therefore equivalent to re-scaling the intensity of choice $\beta$.

The parameter $\Gamma$ in Eqs. (\ref{eq:corr}) measures correlations between the payoff matrix elements $a_{ij}$ and $b_{ji}$. For example if $\Gamma=-1$ one has
\be
{\mathbb E}\left[(a_{ij}+b_{ji})^2\right]={\mathbb E}\left[(a_{ij})^2+(b_{ji})^2+2a_{ij}b_{ji}\right]=0,
\ee
i.e. $a_{ij}=-b_{ji}$ with probability one, corresponding to a zero-sum game. If $\Gamma=1$ one has
\be
{\mathbb E}\left[(a_{ij}-b_{ji})^2\right]=0,
\ee
i.e. $a_{ij}=b_{ji}$ almost surely. For $\Gamma=0$ the payoffs $a_{ij}$ and $b_{ji}$ are uncorrelated. Choices in the interval $\Gamma\in[-1,1]$ interpolate between the extremes. We focus on the regime of anti-correlation, $-1\leq \Gamma\leq 0$ throughout this paper, as we expect this to be more realistic than positively correlated payoffs.
\\

Again, following the spin-glass conventions, and to make sure the thermodynamic limit is well defined, we will re-scale the $\{x_i,y_i\}$ and consider the normalisation $\sum_i x_i=\sum_i y_i=N$. Each of the variables $\{x_i, y_i\}$ is then of order $N^0$. 

At finite $N$ the update rules in discrete time are given by
\BE
x_i(t+1)&=&N\frac{x_i(t)^{1-\alpha}e^{\beta\sum_j a_{ij}y_j(t)}}{\sum_k x_k(t)^{1-\alpha}e^{\beta\sum_j a_{kj}y_j(t)}} \nonumber \\
y_i(t+1)&=&N\frac{y_i(t)^{1-\alpha}e^{\beta\sum_j b_{ij}x_j(t)}}{\sum_k y_k(t)^{1-\alpha}e^{\beta\sum_j b_{kj}x_j(t)}}. \label{eq:mapxy}
\EE
The above choice of scaling now becomes more transparent. The exponentials contain terms of the form $\sum_{j=1}^N a_{ij}y_j$ and $\sum_{j=1}^N b_{ij}x_j$, which are well defined and of order one in the thermodynamic limit ($N\to\infty$) with the above scaling.

In continuous time one has
\BE
\frac{\dot x_i(t)}{x_i(t)}=\left(-r^{-1}\ln x_i(t)+\sum_j a_{ij} y_j(t)-Z_x'(t)\right), \nonumber \\
\frac{\dot y_j(t)}{y_j(t)}=\left(-r^{-1}\ln y_j(t)+\sum_i b_{ji} x_i(t)-Z_y'(t)\right)
\EE
as before, but the Lagrange multipliers are now chosen such that $\sum_ix_i(t)=\sum_i y_i(t)=N$ at all times\footnote{Provided an initial condition fulfilling the normalisation  $\sum_i x_i=\sum_i y_i=N$ is chosen, this can be achieved by setting $Z_x'(t)=N^{-1}\sum_i x_i(t)\left(\sum_j a_{ij}y_j(t)-r^{-1} \ln x_i(t)\right)$ and $Z_y'(t)=N^{-1}\sum_i y_i(t)\left(\sum_j b_{ij}x_j(t)-r^{-1} \ln y_i(t)\right)$.}.
\section{Path-integral analysis}
 \subsubsection{Generating functional description}
We will here describe the technical details of the path-integral analysis of the dynamics. These techniques are standard in the theory of disordered systems, see e.g. \cite{parisimezardvirasoro}, and in particular \cite{coolen,coolen2} for texbook descriptions and a pedagogic review. They have previously been applied to learning in minority game dynamics in \cite{coolen}. The original application to replicator equations is due to Opper and Diederich, see \cite{opper, opper2, opper3}. Other applications of methods from disordered systems to large random games include the calculation of the number of Nash equilibria \cite{berg,berg2}, and the dynamics of random replicator dynamics \cite{galla,galla2}.
\\

The starting point is the continuous dynamics
\BE
\frac{\dot x_i(t)}{x_i(t)}=\left(-r^{-1}\ln x_i(t)+\sum_j a_{ij} y_j(t)-\rho_x(t)+h_{x,i}(t)\right), \nonumber \\
\frac{\dot y_j(t)}{y_j(t)}=\left(-r^{-1}\ln y_j(t)+\sum_i b_{ji} x_i(t)-\rho_y(t)+h_{y,i}(t)\right), \label{eq:dyn}
\EE
where we use the more compact notation $\rho_x(t)$ and $\rho_y(t)$ instead of $Z_x'(t)$ and $Z_y'(t)$. These quantities will be treated as Lagrange multipliers enforcing the normalisation $\sum_i x_i(t)=\sum_j y_j(t)=N$. The fields $h_{i,x}(t)$ and $h_{y,i}(t)$ have been introduced to generate response functions, and will be set to zero at the end of the calculation.
\\
The dynamical generating functional is then given by
\BE\label{eq:gf}
Z[\boldpsi,\boldphi]=\int D[\bx,\by] \delta(\mbox{equations of motion})e^{i\sum_i\int dt \{x_i(t)\psi_i(t)+y_i(t)\varphi_i(t)\}}.
\EE
The source fields $\boldpsi$ and $\boldphi$ have been introduced to generate correlation functions, and will eventually be set to zero at the end of the calculation. The notation $\delta(\mbox{equations of motion})$ indicates that the integral in Eq. (\ref{eq:gf}) is over paths of the dynamics (\ref{eq:dyn}) only, i.e. the delta-functions impose Eqs. (\ref{eq:dyn}) for all $t$ and $i$.

The next step is to write the delta functions in Eq. (\ref{eq:gf}) in their Fourier representation. We then find
\BE
Z[\boldphi,\boldpsi]
&=&\int D[\bx,\by,\widehat\bx,\widehat\by] \exp\Bigg(i\sum_i\int dt \Bigg[\widehat x_i(t)\Bigg(\frac{\dot x_i(t)}{x_i(t)}-\Bigg(-r^{-1}\ln x_i(t)+\sum_j a_{ij} y_j(t)\nonumber\\
&&~~~~~~~~~~~~~~~~~~~~~~~~-\rho_x(t)+h_{x,i}(t)\Bigg)\Bigg)\Bigg]\Bigg)\nonumber \\
&&\times  \exp\left(i\sum_i\int dt \left[\widehat y_i(t)\left(\frac{\dot y_i(t)}{y_i(t)}-\left(-r^{-1}\ln y_i(t)+\sum_j b_{ij} x_i(t)-\rho_y(t)+h_{y,i}(t)\right)\right)\right]\right)\nonumber \\
&&\times \exp\left(i\sum_i\int dt \left[x_i(t)\psi_i(t)+y_i(t)\varphi_i(t)\right]\right).\label{eq:gf0}
\EE
Next, we isolated the terms containing the quenched disorder (the randomly chosen payoff matrix) elements. One has
\BE
Z[\boldphi,\boldpsi]&=&\int D[\bx,\by,\widehat\bx,\widehat\by] \exp\left(i\sum_i\int dt \left[\widehat x_i(t)\left(\frac{\dot x_i(t)}{x_i(t)}+r^{-1}\ln x_i(t)+\rho_x(t)-h_{x,i}(t)\right)\right]\right)\nonumber \\
&&\times  \exp\left(i\sum_i\int dt \left[\widehat y_i(t)\left(\frac{\dot y_i(t)}{y_i(t)}+r^{-1}\ln y_i(t)+\rho_y(t)-h_{y,i}(t)\right)\right]\right)\nonumber \\
&&\times \exp\left(i\sum_i\int dt \left[x_i(t)\psi_i(t)+y_i(t)\varphi_i(t)\right]\right)\nonumber \\
&&\times \exp\left(-i\sum_{ij}\int dt \left[\widehat x_i(t)a_{ij}y_j(t)+\widehat y_j(t) b_{ji}x_i(t)\right]\right).
\EE
We are now in a position to carry out the average over the Gaussian disorder, and to compute ${\mathbb E}[Z[\boldpsi,\boldphi]]$, where ${\mathbb E}[\cdots]$ denotes the disorder-average. We have
\BE
&&{\mathbb E}\left[\exp\left(-i\sum_{ij}\int dt \left[\widehat x_i(t)a_{ij}y_j(t)+\widehat y_j(t) b_{ji}x_i(t)\right]\right)\right] \nonumber \\
&=&\prod_{ij}\exp\Bigg(-\frac{1}{2N}\int dt dt' \{\widehat x_i(t)\widehat x_i(t')y_j(t)y_j(t')+\widehat y_j(t)\widehat y_j(t')x_i(t)x_i(t')\nonumber \\
&&~~~~~~ +\Gamma \widehat x_i(t)x_i(t')y_j(t)\widehat y_j(t')+\Gamma\widehat y_j(t) y_j(t')x_i(t)\widehat x_i(t')\}\Bigg)\nonumber \\
&=&\exp\left(-\frac{1}{2}N\int dt ~dt' \left[L_x(t,t')C_y(t,t')+L_y(t,t')C_x(t,t')+2\Gamma K_x(t,t')K_y(t',t)\right]\right),
\EE
where we have introduced the short-hands
\BE
&C_x(t,t')=\frac{1}{N}\sum_i x_i(t) x_i(t'), ~~~ C_y(t,t')=\frac{1}{N}\sum_j y_j(t) y_j(t'), \nonumber \\
&K_x(t,t')=\frac{1}{N}\sum_i x_i(t) \widehat x_i(t'), ~~~ K_y(t,t')=\frac{1}{N}\sum_j y_j(t) \widehat y_j(t'), \nonumber \\
&L_x(t,t')=\frac{1}{N}\sum_i \widehat x_i(t)\widehat x_i(t'), ~~~ L_y(t,t')=\frac{1}{N}\sum_j \widehat y_j(t)\widehat y_j(t'). \label{eq:op}
\EE
These quantities are introduced into the generating functional by means of delta-functions in their integral representation, e.g.
\BE
1&=&\int D[C_x]\prod_{t,t'}\delta\left(C_x(t,t')-\frac{1}{N}\sum_i x_i(t)x_i(t')\right)\nonumber \\
&=&\int D[\widehat C_x, C]\exp\left(iN\int dt~ dt' \widehat C_x(t,t')\left(C_x(t,t')- N^{-1}\sum_i x_i(t) x_i(t')\right)\right),
\EE
and similarly for the other order parameters in Eq. (\ref{eq:op}). We have chosen the scaling of the conjugate parameter $\widehat C(t,t')$ such that the overall exponent carries a prefactor $N$. 
\\

We then find that the disorder-averaged generating functional can be written in the following form
\be\label{eq:sp}
{\mathbb E}\left[Z[\boldpsi,\boldphi]\right]=\int D[C_x,C_y,L_x,L_y,K_x,K_y,\widehat C_x,\widehat C_y,\widehat L_x, \widehat L_y,\widehat K_x,\widehat K_y]\exp\left(N\left[\Psi+\Phi+\Omega+{\cal O}(N^{-1})\right]\right),
\ee
where
\BE
\Psi&=&i\int dt ~dt' \left[\widehat C_x(t,t')C_x(t,t')+\widehat C_y(t,t')C_y(t,t')+\widehat K_x(t,t')K_x(t,t')+\widehat K_y(t,t')K_y(t,t')\right.\nonumber \\
&&~~~~~~~~~~~~~\left.+\widehat L_x(t,t')L_x(t,t')+\widehat L_y(t,t')L_y(t,t')\right]
\EE
results from the introduction of the above order parameters. The term
\BE
\Phi=-\frac{1}{2}\int dt ~dt' \left[L_x(t,t')C_y(t,t')+L_y(t,t')C_x(t,t')+2\Gamma K_x(t,t')K_y(t',t)\right]
\EE
comes from the disorder average, and $\Omega$ describes the details of the microscopic time evolution
\BE
\Omega&=&N^{-1}\sum_i\log\bigg[\int D[x_i,\widehat x_i] p_{x,0}^{(i)}(x_i(0))\exp\left(i\int dt~ \psi_i(t)x_i(t)\right)\nonumber \\
&&\times \exp\left(i\int dt ~\widehat x_i(t)  \left(\frac{\dot x_i(t)}{x_i(t)}+r^{-1}\ln x_i(t)+\rho_x(t)-h_{x,i}(t)\right)\right)\nonumber \\
&&\times \exp\left(-i\int dt ~ dt' \left[\widehat C_x(t,t')x_i(t)x_i(t')+\widehat L_x(t,t')\widehat x_i(t)\widehat x_i(t')+\widehat K_x(t,t') x_i(t)\widehat x_i(t')\right]\right)\bigg]\nonumber \\
&&+N^{-1}\sum_i\log\bigg[\int D[y_i,\widehat y_i] p_{y,0}^{(i)}(y_i(0))\exp\left(i\int dt ~\varphi_i(t)y_i(t)\right)\nonumber \\
&&\times \exp\left(i\int dt~ \widehat y_i(t)  \left(\frac{\dot y_i(t)}{y_i(t)}+r^{-1}\ln y_i(t)+\rho_y(t)-h_{y,i}(t)\right)\right)\nonumber \\
&&\times \exp\left(-i\int dt ~ dt' \left[\widehat C_y(t,t')y_i(t)y_i(t')+\widehat L_y(t,t')\widehat y_i(t)\widehat y_i(t')+\widehat K_y(t,t') y_i(t)\widehat y_i(t')\right]\right)\bigg]\nonumber \\
\label{eq:omega}
\EE
In this expression $p_{x,0}^{(i)}(\cdot)$ and  $p_{y,0}^{(i)}(\cdot)$ describe the distributions from which initial distributions are drawn. 
\\

The next step is to perform the integrals in Eq. (\ref{eq:sp}) by means of the saddle-point method, valid in the limit $N\to\infty$. This amounts to finding the extrema of the term in the exponent. Setting the variation with respect to the integration variables $C_x,K_x$ and $L_x$ to zero gives
\be
i\widehat C_x(t,t')=\frac{1}{2}L_y(t,t'), ~~~~ i\widehat K_x(t,t')=\Gamma K_y(t',t), ~~~~ i\widehat L_x(t,t')=\frac{1}{2}C_y(t,t'),
\ee
and similarly we obtain
\be
i\widehat C_y(t,t')=\frac{1}{2}L_x(t,t'), ~~~~ i\widehat K_y(t,t')=\Gamma K_x(t',t), ~~~~ i\widehat L_y(t,t')=\frac{1}{2}C_x(t,t')
\ee
from the variation with respect to $C_y,K_y$ and $L_y$.

It remains to perform the extremisation with respect to $\widehat C_x, \widehat K_x,\widehat L_x$, and with respect to the corresponding quantities with subscript $y$. We find
\BE
&C_x(t,t')=\lim_{N\to\infty}N^{-1}\sum_i\avg{x_i(t)x_i(t')}_\Omega, ~~~ C_y(t,t')=\lim_{N\to\infty}N^{-1}\sum_i\avg{y_i(t)y_i(t')}_\Omega, \nonumber \\
&K_x(t,t')=\lim_{N\to\infty}N^{-1}\sum_i\avg{x_i(t)\widehat x_i(t')}_\Omega, ~~~ K_y(t,t')=\lim_{N\to\infty}N^{-1}\sum_i\avg{y_i(t)\widehat y_i(t')}_\Omega, \nonumber \\
&L_x(t,t')=\lim_{N\to\infty}N^{-1}\sum_i\avg{\widehat x_i(t)\widehat x_i(t')}_\Omega, ~~~ L_y(t,t')=\lim_{N\to\infty}N^{-1}\sum_i\avg{\widehat y_i(t)\widehat y_i(t')}_\Omega,
\EE
where the average $\avg{\dots}_\Omega$ is to be taken against a measure defined by the exponent of the expression in Eq. (\ref{eq:omega}), see e.g. \cite{coolen,galla,galla2} for similar calculations. 

Looking back at the definition of the generating functional, Eq. (\ref{eq:gf0}), one also realises that
\BE
C_x(t,t')&=&-\lim_{N\to\infty}N^{-1}\sum_i\left.\frac{\delta ^2 {\mathbb E} \left[Z[\boldpsi,\boldphi]\right]}{\delta \psi_i(t)\delta\psi_i(t')}\right|_{\boldphi=\boldpsi=\bold h=0}, \nonumber \\
K_x(t,t')&=&-\lim_{N\to\infty}N^{-1}\sum_i\left.\frac{\delta ^2 {\mathbb E} \left[Z[\boldpsi,\boldphi]\right]}{\delta \psi_i(t)\delta h_{x,i}(t')}\right|_{\boldphi=\boldpsi=\bold h=0}, \nonumber \\
 L_x(t,t')&=&-\lim_{N\to\infty}N^{-1}\sum_i\left.\frac{\delta ^2 {\mathbb E} \left[Z[\boldpsi,\boldphi]\right]}{\delta h_{x,i}(t)\delta h_{x,i}(t')}\right|_{\boldphi=\boldpsi=\bold h=0},
\EE
and
\BE
C_y(t,t')&=&-\lim_{N\to\infty}N^{-1}\sum_i\left.\frac{\delta ^2 {\mathbb E}\left[Z[\boldpsi,\boldphi]\right]}{\delta \varphi_i(t)\delta\varphi_i(t')}\right|_{\boldphi=\boldpsi=\bold h=0}, \nonumber \\
K_y(t,t')&=&-\lim_{N\to\infty}N^{-1}\sum_i\left.\frac{\delta ^2 {\mathbb E}\left[Z[\boldpsi,\boldphi]\right]}{\delta \varphi_i(t)\delta h_{y,i}(t')}\right|_{\boldphi=\boldpsi=\bold h=0}, \nonumber \\
 L_y(t,t')&=&-\lim_{N\to\infty}N^{-1}\sum_i\left.\frac{\delta ^2 {\mathbb E}\left[Z[\boldpsi,\boldphi]\right]}{\delta h_{y,i}(t)\delta h_{y,i}(t')}\right|_{\boldphi=\boldpsi=\bold h=0}.
\EE
Given that $Z[\boldpsi=0,\boldphi=0,\bh]=1$ for all $\bh$ due to normalisation we conclude that $L_x(t,t')=L_y(t,t')=0$ for all $t,t'$.

The variables $\boldpsi$ and $\boldphi$ have now served their purpose (to generate correlation functions), and we set them to zero. We will also assume uniform perturbations $h_{i,x}(t)\equiv h_x(t)$ and $h_{y,j}(t)=h_y(t)$ for all $i$, and that initial conditions are chosen from identical distributions for all components $x_i$ and $y_i$ (i.e. $p_{x,0}^{(i)}(\cdot)$ does not depend on $i$, and similarly for $p_{y,0}^{(i)}(\cdot)$. Then we have
\BE
\Omega&=&\log\bigg[\int D[x,\widehat x] p_{x,0}(x(0)) \exp\left(i\int dt ~\widehat x(t)  \left(\frac{\dot x(t)}{x(t)}+r^{-1}\ln x(t)+\rho_x(t)-h_x(t)\right)\right)\nonumber \\
&&\times \exp\left(-\int dt ~ dt' \left[ \frac{1}{2}C_y(t,t')\widehat x(t)\widehat x(t')+i\Gamma G_y(t',t) x(t)\widehat x(t')\right]\right)\bigg]\nonumber \\
&&+\log\bigg[\int D[y,\widehat y] p_{y,0}(y(0) \exp\left(i\int dt~ \widehat y(t)  \left(\frac{\dot y(t)}{y(t)}+r^{-1}\ln y(t)+\rho_y(t)-h_y(t)\right)\right)\nonumber \\
&&\times \exp\left(-\int dt ~ dt' \left[\frac{1}{2}C_x(t,t')\widehat y(t)\widehat y(t')+i\Gamma G_x(t,t') y(t)\widehat y(t') \right]\right)\bigg],\label{eq:omega2}
\EE
where we have used the above saddle-point results, and where we have introduced $G_x(t,t')=-iK_x(t,t')$ and $G_y(t,t')=-iK_y(t,t')$.

The resulting term
\BE
Z_{\mbox{\footnotesize eff}}&=&\int D[x,\widehat x] D[y,\widehat y]p_{x,0}(x(0)) p_{y,0}(y(0) \exp\left(i\int dt ~\widehat x(t)  \left(\frac{\dot x(t)}{x(t)}+r^{-1}\ln x(t)+\rho_x(t)-h_x(t)\right)\right)\nonumber \\
&&\times \exp\left(-\int dt ~ dt' \left[ \frac{1}{2}C_y(t,t')\widehat x(t)\widehat x(t')+i\Gamma G_y(t',t) x(t)\widehat x(t')\right]\right)\nonumber \\
&&\times \exp\left(i\int dt~ \widehat y(t)  \left(\frac{\dot y(t)}{y(t)}+r^{-1}\ln y(t)+\rho_y(t)-h_y(t)\right)\right)\nonumber \\
&&\times \exp\left(-\int dt ~ dt' \left[\frac{1}{2}C_x(t,t')\widehat y(t)\widehat y(t')+i\Gamma G_x(t,t') y(t)\widehat y(t') \right]\right)
\EE
is recognised as the generating function of the {\em effective} dynamics
\BE
\dot x(t)=x(t)\left[\Gamma\int dt' G_y(t,t') x(t')-r^{-1}\ln x(t)-\rho_x(t)+\eta_x(t)+h_x(t)\right] \nonumber \\
\dot y(t)=y(t)\left[\Gamma\int dt' G_x(t,t') y(t')-r^{-1}\ln y(t)-\rho_y(t)+\eta_y(t)+h_y(t)\right], \label{eq:effective}
\EE
where
\BE
 &G_x(t,t')=\avg{\frac{\delta x(t)}{\delta h_x(t')}}_*,~~~~ G_y(t,t')=\avg{\frac{\delta y(t)}{\delta h_y(t')}}_*, \nonumber \\
 &\avg{\eta_x(t)\eta_x(t')}_*=\avg{y(t)y(t')}_*,~~~~\avg{\eta_y(t)\eta_y(t')}_*=\avg{x(t)x(t')}_*, \nonumber \\
 &\avg{x(t)}_*=\avg{y(t)}_*=1,
\EE
and where $\avg{\cdots}_*$ denotes an average over realizations of the effective dynamics (\ref{eq:effective}).
This is to be evaluated at vanishing perturbation fields $h_x(t)=h_y(t)=0$. It is hence appropriate to consider
\BE
\dot x(t)=x(t)\left[\Gamma\int dt' G_y(t,t') x(t')-r^{-1}\ln x(t)-\rho_x(t)+\eta_x(t)\right], \nonumber \\
\dot y(t)=y(t)\left[\Gamma\int dt' G_x(t,t') y(t')-r^{-1}\ln y(t)-\rho_y(t)+\eta_y(t)\right], \label{eq:effcont}
\EE
where\footnote{The constraints $\avg{x(t)}_*=\avg{y(t)}_*$ are here a reflection of the normalisation $\sum_i x_i(t)=\sum_j y_j(t)=N$ in the microscopic model. They can formally be derived by introducing an delta-function in the original generating functional, imposing the microscopic constraint. This has been omitted here to reduce the overall complexity of the calculation. Similar methods have been used e.g. in \cite{spherical}.}
\BE
 &G_x(t,t')=\avg{\frac{\delta x(t)}{\delta \eta_x(t')}}_*,~~~~ G_y(t,t')=\avg{\frac{\delta y(t)}{\delta \eta_y(t')}}_*, \nonumber \\
 &\avg{\eta_x(t)\eta_x(t')}_*=C_y(t,t')=\avg{y(t)y(t')}_*,~~~~\avg{\eta_y(t)\eta_y(t')}_*=C_x(t,t')=\avg{x(t)x(t')}_*, \nonumber \\
 &\avg{x(t)}_*=\avg{y(t)}_*=1. \label{eq:sc}
\EE
We note that the path-integral analysis up to this point can also be carried out for the discrete dynamics. In this case one obtains the following effective process:
 \BE
 x(t+1)&=&\frac{x(t)^{1-\alpha}\exp\left(\beta\left[\Gamma\sum_{t'} G_y(t,t')x(t')+\eta_x(t)\right]\right)}{Z_x(t)}\nonumber \\
 y(t+1)&=&\frac{y(t)^{1-\alpha}\exp\left(\beta\left[\Gamma\sum_{t'} G_x(t,t')y(t')+\eta_y(t)\right]\right)}{Z_y(t)}, \label{eq:effmap}
 \EE
with self-consistency relations as in Eq. (\ref{eq:sc}). Due to causality we have $G(t,t')=0$ for $t'\geq 0$, both in the continuous-time and in the discrete-time case, so the integrals over $t'$ in Eqs. (\ref{eq:effcont}) and the sums in Eq. (\ref{eq:effmap}) only extend over the range $t'<t$.

 \subsubsection{Fixed point analysis}
In the stationary state all two time quantities (e.g. $C_x(t,t'), G_x(t,t')$) become functions of time differences only, i.e. $G_x(t,t')=G_x(\tau)$, where $\tau=t-t'$, and similar for the other two-time observables. Assuming the dynamics reaches a fixed point one also has $C_x(t,t')\equiv \mbox{const}$ and similarly for $C_y(t,t')$.

Fixed points of the discrete-time effective dynamics (\ref{eq:effmap}) are given by  
\BE
-\alpha\ln x^*+\Gamma\beta\chi_y x^*+\beta \eta_x^*-\ln Z_x^*&=&0,\nonumber \\
-\alpha\ln y^*+\Gamma\beta\chi_x y^*+\beta \eta_y^*-\ln Z_y^*&=&0,
\EE
where we have written $\chi_x=\int_0^\infty d\tau~ G_x(\tau)$ and $\chi_y=\int_0^\infty d\tau ~G_y(\tau)$. An asterisk as a superscript indicates fixed-point quantities as before. From the continuous-time effective process, Eq. (\ref{eq:effcont}), one obtains the equivalent fixed-point condition
\BE
-r^{-1}\ln x^*+\Gamma\chi_y x^*+\eta_x^*-\rho_x^*&=&0,\nonumber \\
-r^{-1}\ln y^*+\Gamma\chi_x y^*+\eta_y^*-\rho_y^*&=&0.
\EE
Due to symmetry we expect $\chi_x=\chi_y\equiv \chi$, $\rho_x^*=\rho_y^*\equiv\rho$, see also \cite{berg,berg2}. We will also write
\be
q\equiv \avg{(x^*)^2}_*=\avg{(y^*)^2}_*.
\ee
Let us write $\eta_x=\sqrt{q}z$ with $z$ a static Gaussian random variable of mean zero and unit variance. Then let $x(z)$ be the positive solution, $x$, of
\be
-r^{-1}\ln x+\Gamma\chi x+\sqrt{q}z-\rho=0.\label{eq:fpeff}
\ee
The order parameters $\chi$, $q$ and $\rho$ are to be determined from the self-consistency relations
\be
\chi=\frac{1}{\sqrt{q}}\avg{\frac{\partial x(z)}{\partial z}}_*, ~~ q=\avg{(x(z))^2}_*, ~~ \avg{x(z)}_*=1,
\ee
in other words, we have
\BE
\chi&=&\frac{1}{\sqrt{q}}\int_{-\infty}^\infty Dz ~\frac{\partial x(z)}{\partial z}, \nonumber \\
q&=&\int_{-\infty}^\infty Dz~ x(z)^2, \nonumber \\
1&=&\int_{-\infty}^\infty Dz~x(z),\label{eq:fpsc}
\EE
where $Dz=\frac{dz}{\sqrt{2\pi}}e^{-z^2/2}$. These equations fully determine the statistical properties of the fixed points of the dynamics, and can be used to compute quantities such as the distribution of frequencies with which pure actions are played (i.e. the shape of the resulting mixed strategy profile), or the entropy of mixed strategies. Theoretical predictions are tested against simulations below (see Sec. \ref{sec:sim}).

\subsubsection{Linear stability analysis}
We will now carry out a linear stability analysis of the effective dynamics in the continuous-time case. We mostly follow the approach first proposed in \cite{opper,opper3}. As a first step we assume the dynamics is perturbed by small noise terms, $\xi(t)$ and $\zeta(t)$:
\BE
\dot x(t)=x(t)\left[\Gamma\int dt' G_y(t,t') x(t')-r^{-1}\ln x(t)-\rho_x(t)+\eta_x(t)+\xi(t)\right], \nonumber \\
\dot y(t)=y(t)\left[\Gamma\int dt' G_x(t,t') y(t')-r^{-1}\ln y(t)-\rho_y(t)+\eta_y(t)+\zeta(t)\right].
\EE
and that we have small perturbation about a fixed point, i.e.
\BE
x(t)&=&x^*+\widehat x(t),\\
y(t)&=&y^*+\widehat y(t),\\
\eta_x(t)&=&\eta_x^*+\widehat v(t), \\
\eta_y(t)&=&\eta_y^*+\widehat w(t).
\EE
Perturbations are here labelled by hats on the corresponding variables, this is not to be confused with the notation $\widehat x_i, \widehat y_j$ etc in earlier sections, where, in the course of computing the generating functional, hats indicated conjugate variables. Following \cite{opper,opper2} we restrict the analysis to cases where $x^*>0$ and $y^*>0$. Expanding to linear order in the deviations from the fixed point we then have
\BE
\frac{d}{dt}\widehat x(t)=-r^{-1}\widehat x(t)+x^*\left[\Gamma\int dt'~ G_y(t-t')\widehat x(t')+\widehat v(t)+\xi(t)\right], \nonumber \\
\frac{d}{dt}\widehat y(t)=-r^{-1}\widehat y(t)+y^*\left[\Gamma\int dt'~ G_x(t-t')\widehat y(t')+\widehat w(t)+\zeta(t)\right]. 
\EE
In Fourier space we have
\BE
\left[\frac{i\omega+r^{-1}}{x^*}-\Gamma\widetilde G_y(\omega)\right]\widetilde x(\omega)=\widetilde v(\omega)+\widetilde\xi(\omega), \nonumber \\
\left[\frac{i\omega+r^{-1}}{y^*}-\Gamma\widetilde G_x(\omega)\right]\widetilde y(\omega)=\widetilde w(\omega)+\widetilde\zeta(\omega), 
\EE
for which we will introduce the short-hand notation
\BE
A(\omega,x^*)\widetilde x(\omega)=\widetilde v(\omega)+\widetilde\xi(\omega), \nonumber \\
B(\omega,y^*)\widetilde y(\omega)=\widetilde w(\omega)+\widetilde\zeta(\omega).
\EE
Denoting the fraction of strategies played with non-zero probability by $\phi$ (not to be confused with the memory-loss parameter $\phi$ in earlier sections), and taking into account that we are only considering components with $x^*>0, y^*>0$  this gives (for details of similar calculations see \cite{opper})
\BE
\avg{|\widetilde x(\omega)|^2}_*&=&\phi\left(\avg{|\widetilde y(\omega)|^2}_*+1\right) \avg{\frac{1}{|A(\omega,x^*)|^2}}_*, \nonumber \\
\avg{|\widetilde y(\omega)|^2}_*&=&\phi\left(\avg{|\widetilde x(\omega)|^2}_*+1\right) \avg{\frac{1}{|B(\omega,x^*)|^2}}_* \label{eq:inst1},
\EE
where we have used the self-consistency relations $\avg{|\widetilde v(\omega)|^2}_*=\avg{|\widetilde y(\omega)|^2}_*$ and $\avg{|\widetilde w(\omega)|^2}_*=\avg{|\widetilde x(\omega)|^2}_*$.

Again following \cite{opper} let us now focus on the $\omega=0$ mode. Using the symmetry between players we have $\avg{|\widetilde x(\omega=0)|^2}_*=\avg{|\widetilde y(\omega=0)|^2}_*$, and hence we find
\be\label{eq:fluct}
\avg{|\widetilde x(\omega=0)|^2}_*=\left[\frac{1}{\phi \avg{\frac{1}{|A(\omega=0,x^*)|^2}}_*}-1\right]^{-1}.
\ee
This expression diverges, as 
\be\label{eq:instab}
\phi \avg{\frac{1}{|\frac{r^{-1}}{x^*}-\Gamma\chi|^2}}_*=1,
\ee
signalling the onset of instability. In particular Eq. (\ref{eq:fluct}) predicts a negative value of $\avg{|\widetilde x(\omega=0)|^2}_*$, if $\phi \avg{\frac{1}{|A(\omega=0,x^*)|^2}}_*<1$, indicating that our self-consistent fixed-point solution breaks down. Eq. (\ref{eq:instab}) therefore defines the boundary of the stable fixed point phase, and was used to generate the stability diagram in the main paper (Fig. 2). The fraction of active strategies is here given by $\phi=1$, following our solution for fixed points of the effective process (we find that Eq. (\ref{eq:fpeff}) has positive solutions $x(z)$ for all values of $z$, provided $\Gamma<0$).
\section{Numerical methods and simulation results}

\begin{figure}[t]
\vspace{-1em}
\centerline{\includegraphics[width=0.4\textwidth]{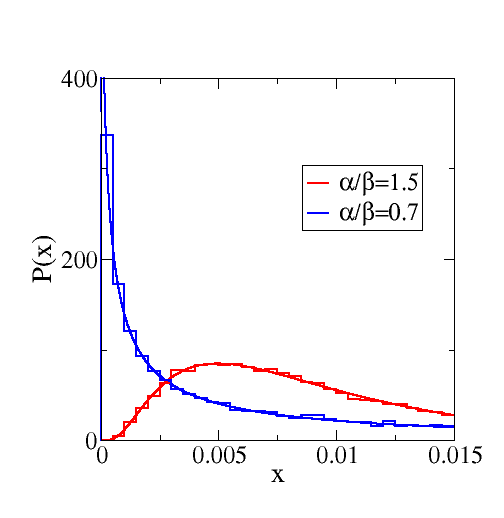}\includegraphics[width=0.4\textwidth]{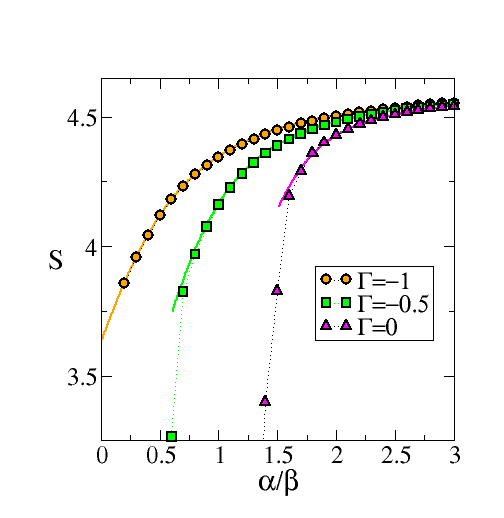}}

\caption{Test of theoretical predictions for the stable phase against simulations. The left-hand panel shows the distribution of components $x_i$ of mixed strategies at the fixed point ($\Gamma=-0.5$). Solid lines are theoretical predictions, noisy lines from simulations. The right-hand panel shows the entropy $S$ of the fixed point strategies of players. Symbols are from simulations (at $N=100$ strategies per player, simulations run for $7,500$ time steps, with measurements starting after $5,000$ time steps). Averages over $100$ different payoff matrices are taken. Solid lines are from the theory, hence only shown in the stable phase in the right-hand panel. Agreement with simulations is good, except for small deviations near the onset of instability. We attribute these to finite-size and equilibration effects. All data in this figure is taken at $\beta=0.01$.}
\label{fig:suppfig1}
\end{figure}

\subsection{Test of theoretical predictions against simulations}\label{sec:sim}
\subsubsection{Order parameters in fixed point phase}
Eqs. (\ref{eq:fpsc}) together with Eq. (\ref{eq:fpeff}) are the final result of our path-integral analysis in the fixed-point phase. These equations determine the relevant order parameters $\chi, q$ and $\rho$ self-consistently. We notice the high degree of nonlinearity due to the logarithmic term in (\ref{eq:fpeff}). In absence of this term (i.e. for $r^{-1}=0$) the resulting equations are linear and the Gaussian integrals in (\ref{eq:fpsc}) can be carried out and the resulting equations can be simplied further, see \cite{opper,galla,galla2}) for details. In the presence of memory-loss ($r^{-1}>0$) this is not possible however, and we have to approach the self-consistency problem numerically. We here restrict the analysis to the case $\Gamma<0$, when a positive solution of (\ref{eq:fpeff}) is found for all values of $z$. Numerically solving Eq. (\ref{eq:fpeff}) gives $x(z)$ with an iterative Newton-Raphson procedure then allows us to determine the order parameters $\chi,q,\rho$ \footnote{The integrals in Eq. (\ref{eq:fpsc}) are evaluated numerically, and the integration range necessarily needs to be truncated during this procedure. Our results are therefore numerical estimates of the actual solution.}. Once these order parameters are determined the distribution of the components of the strategy vectors can be obtained from solving the above Eq. (\ref{eq:fpeff})
\begin{displaymath}
-r^{-1}\ln x(z)+\Gamma\chi x(z)+\sqrt{q}z-\rho=0.
\end{displaymath}
More precisely one has
\be
P(x)=\int dz \frac{e^{-z^2/2}}{\sqrt{2\pi}}~ \delta(x-x(z))\label{eq:pofz}
\ee
for the distribution of fixed points of the effective process. Recalling that degrees of freedom in the path-integral analysis have been obtained from the original strategy components by a re-scaling with a factor of $N$ ($\sum_i x_i=N$ instead of $\sum_i x_i=1$), an analytical prediction for the  distribution of strategy components of the original problem at a large but finite value of $N$ can be obtained using Eq. (\ref{eq:pofz}), and upon undoing this re-scaling. Results are shown in Fig. \ref{fig:suppfig1} of this Supplementary Information (left-hand panel). As seen in the figure the analytical predictions for this highly non-trivial and non-Gaussian distribution agree rather well with results from direct simulations of the original learning dynamics.
\\

We can also determine the entropy of a typical mixed strategy of a system at finite $N$ at the fixed point as follows. Given the normalisation $\sum_i x_i=N$ we define $S$ to be the entropy of the mixed strategy vector $(x_1/N,\dots,x_N/N)$, i.e.
\BE
S&=&-\sum_i \frac{x_i}{N}\ln\frac{x_i}{N}\nonumber \\
&=&-\frac{1}{N}\sum_i x_i\ln x_i +\ln(N) \nonumber \\
&=&-\avg{x(z)\ln(x(z))}_z+\ln(N),
\EE
where $\avg{\dots}_z$ denotes an average over $z$, i.e. $\avg{\cdots}_z=\int dz \cdots \frac{e^{-z^2/2}}{\sqrt{2\pi}}$. Results are shown in Fig. \ref{fig:suppfig1} of this Supplementary Information (right panel), and again theoretical predictions and direct measurements from simulations agree very well. We note that mixed strategies concentrate on the centre of strategy space the for $\alpha/\beta\to\infty$, i.e. for very quick memory loss. In this case one has $x_i\approx1$ for all $i$ (recall the normalisation $\sum_i x_i=N$), i.e.
\be
S=-\sum_i \frac{x_i}{N}\ln\frac{x_i}{N}\approx -\sum_i \frac{1}{N}\ln\frac{1}{N}=\ln(N).
\ee
As a final remark we point out that Eqs. (\ref{eq:fpsc}) and Eq. (\ref{eq:fpeff}) are valid only in the fixed point phase, as the assumption of a fixed point was explicitly made in deriving these relations. We are therefore only able to predict the statistics of the solution in the stable fixed point phase. The solution of the effective dynamics below the transition, in the chaotic regime, is a formidable task. No promising approaches are available, similar to lack of analytical handles for example on the `turbulent' so-called non-ergodic phase of the minority game \cite{coolen}.
\subsubsection{Onset of instability}
\begin{figure}[t]
\vspace{-2em}
\centerline{\includegraphics[width=0.5\textwidth]{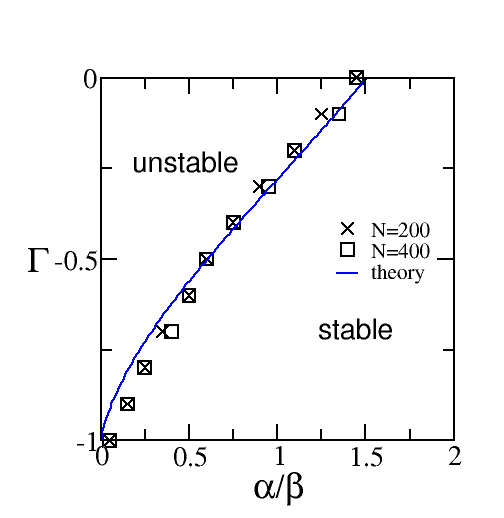}}
\caption{Test of theoretical predictions for the stability diagram. Solid line shows the onset of instability as predicted by the theory (see Eq. (\ref{eq:instab})). Markers show results from simulations (see text for details). All data in this figure is taken at $\beta=0.01$.}
\label{fig:suppfig2}
\end{figure}
The validity of the analytical predictions for the onset of instability (boundary of the chaotic phase) has already been successfully confirmed in simulations in Fig. 2 of the main paper, where we have measured the expected dimension of the dynamical attractors in parameter space. These simulations are time-consuming and were therefore limited to systems of dimension $2(N-1)=98$. In order to provide a more precise verification we have determined the onset of instability in larger systems in Fig. \ref{fig:suppfig2} of this Supplementary Material. The numerical data is here obtained as follows:
\begin{itemize}
\item[1.] For a fixed value of $\Gamma$ generate $M$ samples of the payoff bi-matrix.
\item[2.] For these $M$ realisations of the game, run the dynamics at large $\alpha/\beta$ and, for each sample determine whether or not it reaches a stable fixed point.
\item[3.] If the majority of the $M$ samples converges to a fixed point, lower the value of $\alpha/\beta$ and repeat step 2 until more than half of the samples no longer converge.
\item[4.] Record this value of $\alpha/\beta$ as the onset of instability, and proceed to a new value of $\Gamma$ in 1.
\end{itemize}
In the simulations of Fig. \ref{fig:suppfig2} we have used $M=10$ samples. A given run is considered to reach a fixed point if both  (i) all eigenvalues of the Jacobian at a final time $T$ are within the unit circle and (ii) the total fluctuations $N^{-1}\sum_i\left[3/T\sum_{t=2/3T}^T x_i(t)^2-\left(3/T\sum_{t=2/3T}^T x_i(t)\right)^2\right]$ are less that a pre-defined threshold $\vartheta$. In our simulations we have used $T=15,000$ and $\vartheta=10^{-5}$. If these criteria are not fullfilled the run is considered not to converge. We cannot entirely exclude to identify runs as non-convergent, when in fact they do converge on time scales larger than $T$. In this sense we can not exclude a potential over-estimation of the value $\alpha/\beta$ at which the instability sets in in the numerical results presented in Fig. \ref{fig:suppfig2}. The agreement with the theoretical predictions is very good however. Small deviations can be attributed to the effect just discussed, and to the fact that the theoretical prediction of the instability line is obtained for the continuous-time dynamics, whereas simulations are carried out for the discrete-time map at $\beta=0.01$. Additionally there may be potential finite size effects.

\subsection{Estimation of the attractor dimension}
\begin{figure}[t]
\vspace{-0em}
\centerline{\includegraphics[width=0.65\textwidth]{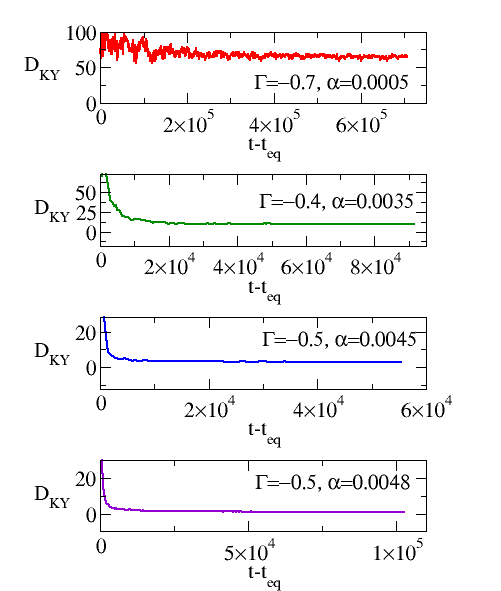}}
\caption{Convergence of measurements of attractor dimensions. Data is from simulations of games with $N=50$ strategies per player, and corresponds to the four attractors shown in Fig.1 of the main manuscript.}
\label{fig:suppfig3}
\end{figure}

The Liapunov spectrum of the attractors are determined using a procedure similar to that described in \cite{sandri}.  Measurements are started after some equilibration time ($t_{\mbox{eq}}=150,000$ iterations), after which we run a linearised map
\be
\bz(t+1)=J_{\bx(t),\by(t)}\bz(t),
\ee
parallel to the simulation of the original system, with $L=2(N-1)$ degrees of freedom. The $L\times L$ matrix $J_{\bx(t),\by(t)}$ is the Jacobian of the full non-linear system. We run $L$ copies of the linearized dynamics, $z^{(1)},\dots,z^{(L)}$ started from the $L$ unit vectors. We then regularly perform a stabilized Gram-Schmidt procedure, and obtain estimates of the Liapunov exponents \cite{sandri}. From these estimates one then calculates the Kaplan-Yorke dimension as
\be
D=j-\frac{\sum_{i=1}^j \lambda_i}{\lambda_{j+1}},
\ee
where the Liapunov exponents are ordered as $\lambda_1\geq \lambda_2\geq\dots\geq\lambda_L$, and where $j$ is the largest integer such that $\lambda_1+\dots+\lambda_j\geq0$ \cite{kaplan,sandri}.
The estimates of the attractor dimension may fluctuate as the simulation run continues after equilibration, and as the attractor is sampled. In practice we find that the measured dimension tends to converge in most runs as the duration of the simulation increases. In our simulations we consider the attractor dimension in a given run as converged when the difference between the maximum and minimum estimate of the dimension in a time window of $20,000$ iteration steps deviate by less than $5\%$ from each other. The dimension reported is then the average over that time window. In other words, simulations are first run for $150,000$ steps to equilibrate, then at least $20,000$ iterations additional are performed during which measurements are taken.  Subsequently the simulation is extended (up to at most $10^6$ iterations) until the convergence criterion is met. In practice we find that most samples have converged at $10^6$ iterations or earlier, when we terminate our simulation. Examples of such measurements are shown in Fig. \ref{fig:suppfig3} of this Supplementary Information, the data shown corresponds to the attractors shown in Fig. 1 of the main paper. Samples that have not converged on this time scale are ignored in our analysis, and have been disregarded when compiling the data for Fig. 2 of the main paper.  We here find that only a small fraction of samples converges when the attractor dimension is very high.

\subsection{Return distribution}
\begin{figure}[t]
\vspace{-0em}
\centerline{\includegraphics[width=0.5\textwidth]{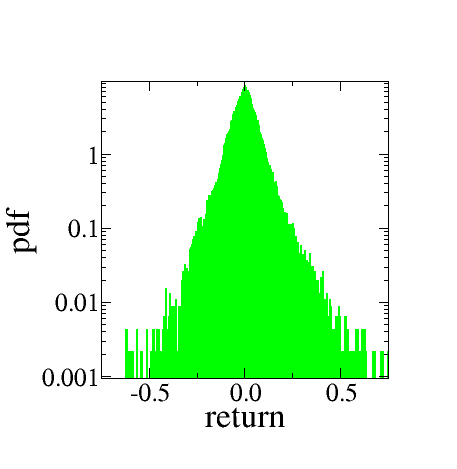}}
\caption{Distribution of returns for the run shown in Fig. 3 of the main manuscript.}
\label{fig:suppfig4}
\end{figure}

The time series of `returns' in Fig. 3 of the main manuscript shows the changes of {\em total} payoff to the two players. Specifically, we measure
\be
\Pi_{\mbox{{\small tot}}}(t)=\sum_i\sum_j  \left\{x_i(t) a_{ij} y_j(t)+y_i(t) b_{ij} x_j(t)\right\} 
\ee
at each time step $t$ in the equilibrated regime, and then plot $\Pi_{\mbox{{\small tot}}}(t)-\Pi_{\mbox{{\small tot}}}(t-1)$ in Fig. 3 of the main paper. The corresponding distribution of returns is shown in Fig. \ref{fig:suppfig4} of this SI, and shows exponential tails.



\begin{thebibliography}{99}
\bibitem{nash} J. Nash, Equilibrium points in n-person games, Proceedings of the National Academy of Sciences {\bf 36} (1) 48-49 (1950)
\bibitem{vonneumann} J. von Neumann, O. Morgenstern, {Theory of Games and Economic Behaviour}, Princeton University Press, Princeton NJ (2007)
\bibitem{berg} A. McLennan, J. Berg, The asymptotic expected number of Nash equilibria of two player normal form games, Games and Economic Behavior 51(2), 264-295 (2005)
\bibitem{berg2}  J. Berg, M. Weigt, Entropy and typical properties of Nash equilibria in two-player Games, Europhys. Lett. 48(2), 129-135 (1999).
\bibitem{opper} M. Opper, S. Diederich, Phase transition and $1/f$ noise in a game dynamical model, Phys. Rev. Lett. {\bf 69} 1616-1619 (1992)
\bibitem{opper2} S. Diederich, M. Opper, Replicators with random interactions: A solvable model, Phys. Rev. A {\bf 39} 4333-4336 (1989).
\bibitem{ho} T. H. Ho, C. F. Camerer, J.-K. Chong, Self-tuning experience weighed attraction learning in games, J. Econ. Theor. {\bf 133} 177-198 (2007)
\bibitem{camerer1} C. Camerer, T.H. Ho, Experience-weighted attraction learning in normal form games, Econometrica {\bf 67} (1999) 827
\bibitem{camerer} C. Camerer, Behavioral Game Theory: Experiments in Strategic Interaction (The Roundtable Series in Behavioral Economics), Princeton University Press, Princeton NJ, 2003
\bibitem{fudenberg} D. Fudenberg, D.K. Levine, {\em Theory of Learning in Games}, MIT Press, Cambridge MA (1998)
\bibitem{young} H. P. Young, {\em Individual Strategy and Social Structure: An Evolutionary Theory of Institutions}, Princeton University Press, Princeton NJ (1998)
\bibitem{hommes} W. A. Brock, C. H. Hommes, Heterogeneous beliefs and routes to chaos in a simple asset pricing model, J. Econ. Dyn. and Contr. {\bf 22} 1235-1274 (1998)
\bibitem{skyrms} B. Skyrms, Chaos in game dynamics, J. of Logic, Language and Information {\bf 1} 111-130 (1992)
\bibitem{sato} Y. Sato, E. Akiyama, J. D. Farmer, Chaos in learning a simple two-player game, Proc. Nat. Acad. Sci. USA {\bf 99} 4748-4751 (2002)


\bibitem{sato2} Y. Sato, J.-P. Crutchfield, Coupled replicator equations for the dynamics of learning in multiagent systems, Phys. Rev. E {\bf 67} 015206(R) (2003)

\bibitem{nowak} M. A. Nowak, {\em Evolutionary dynamics}, Harvard University Press, Cambridge MA (2006)
\bibitem{sigmund} J. Hofbauer, K. Sigmund, {\em Evolutionary games and population dynamics}, Cambridge University Press, Cambridge, 1998
\bibitem{may} R. M. May, Will a Large Complex System be Stable? Nature {\bf 238}, 413 - 414 (1972); 
\bibitem{f1} Note that the fixed point reached in the stable regime is only a Nash equilibrium at $\Gamma=0$ and in the limit $\alpha\to 0$. When $\alpha > 0$ the players are effectively assuming their opponent's behavior is non-stationary, and that more recent moves are more useful than moves in the distant past.
\bibitem{dedominicis} De~Dominicis, C., Phys. Rev. B  {\bf 18} 4913-4919 (1978)
\bibitem{ghashghaie} Ghashghaie, S., Breymann, W., Peinke, J., Talkner, P., Dodge, Y., Turbulent cascades in foreign exchange markets, Nature {\bf 381} 767-770 (1996)
\bibitem{f2} In contrast to financial markets, for the behavior we observe here the distribution of heavy tails decay exponentially (as opposed to following a power law).  We hypothesize that this is because the players in financial markets use a variety of different timescales $\alpha$.
\bibitem{lorenz} Lorenz, E. N., Atmospheric predictability revealed by naturally occurring analogues, J. Atmos. Sci. {\bf 26}, 636-646 (1969)
\bibitem{farmer} J. D. Farmer, J. J. Sidorowich, Predicting chaotic time series, Phys. Rev. Lett. {\bf 59} 845-848 (1987)




\bibitem{camerer2} T . H. Ho, C. F. Camerer, J.-K. Chong, {\em Self-tuning experience weighted attraction learning in games}, J. Econ. Theory {\bf 133} (2007) 177

\bibitem{satofarmer} Y. Sato. D. Farmer, in preparation
\bibitem{gallaprl} T. Galla, {\em Intrinsic noise in game dynamical learning}, Phys. Rev. Lett. {\bf 103} (2009) 198702
\bibitem{realpe} J. Realpe-Gomez J. et al., {\em Fixation and escape times in stochastic game learning}, submitted (2011), preprint available at {\tt http://arxiv.org/abs/1102.0876}
\bibitem{gallanoise} T. Galla, {\em Cycles between cooperation and defection in imperfect learning}, submitted 2011, preprint available at {\tt http://arxiv.org/abs/1101.4378}
\bibitem{vilone} D. Vilone, A. Robledo, A. Sanchez, Chaos and unpredictability in evolutionary games in discrete time, preprint {\tt http://arxiv.org/abs/1103.1484}
\bibitem{parisimezardvirasoro} G. Parisi, M. Mezard, M. A. Virasoro, {\em Spin glass theory and beyond}, World Scientific Publishing, Singapore (1987)
\bibitem{coolen2} A. C. C. Coolen, in Handbook of Biological Physics Vol 4 (Elsevier Science 2001; eds. F. Moss and S. Gielen), 597-662 
ÔStatistical mechanics of Recurrent Neural networks II: DynamicsÕ 
\bibitem{coolen} A. C. C. Coolen, {\em The mathematical theory of minority games}, Oxford University Press, Oxford UK (2005)
\bibitem{opper3} M. Opper, S. Diederich, Replicator Dynamics, Computer Physics Communications Volumes 121-122, September-October 1999, Pages 141-144
Proceedings of the Europhysics Conference on Computational Physics CCP 1998 
\bibitem{galla} T. Galla, Random replicators with asymmetric couplings, J. Phys. A: Math. and Gen. {\bf 39} 3853 (2006)
\bibitem{galla2} T. Galla, Two-population replicator dynamics and number of Nash equilibria in matrix games, EPL (Europhysics Letters) {\bf 78} 20005 ( 2007)
\bibitem{spherical} T. Galla, A. C. C. Coolen, D. Sherrington, Dynamics of a spherical minority game, J . Phys. A: Math. Gen. 36 (2003) 11159-11172
\bibitem{sandri} M. Sandri, Numerical calculation of Lyapunov exponents, The Mathematica Journal {\bf 6}, 78-84 (1996)
\bibitem{kaplan} J. Kaplan, J. A. Yorke, In {\em Functional Differential Equations and Approximations of Fixed Points: Proceedings}, Bonn, July 1978 (Ed. H.-O. Peitgen and H.-O. Walther). Berlin: Springer-Verlag, p. 204, 1979.
 

\end{thebibliography}
\end{document}